# The monochromatic X-rays facilities at NIM

Guo Siming[1], Wu Jinjie[1*], Hou Dongjie[1,2], Zhou Pengyue[1], Wang Eryan[1], Song Ruiqiang[1], Wang Jia[1], Zhai Yudan[1], Liu Haoran[1], Li Xinqiao[2], An Zhenghua[2], Zhang Dali[2], Peng Wenxi[2], Zhou Xu[2], Li Mengshi[1], Li Chengze[1], Zhang Shuai[1], Ren Guoyue[1], Wang Ji[1], Huang Jianwei[1], Li Dehong[1], Zhang Jian[1]

1. National Institute of Metrology, Beijing, China
2. Institute of High Energy Physics, CAS, Beijing, China

**Abstract,** Space scientific exploration is becoming the main battlefield for mankind to explore the universe. Countries around the world have successively launched various space exploration satellites. Accurate calibration on the ground is a key factor for space science satellites to obtain observational results. In order to provide calibration for various satellite-borne detectors, several monochromatic X-rays facilities has been built at National Institute of Metrology, P.R. China (NIM). These facilities are mainly based on grating diffraction and Bragg diffraction, the energy range of produced monochromatic X-rays is (0.218-301) keV. The facilities have a good performance on energy stability, monochromaticity and flux stability. Monochromaticity of all facilities is better than 3.0%, the stability of energy is better than 1.0% over 8 hours, and the stability of flux is better than 2.0% over 8 hours. The calibration experiments of satellite-borne detectors, such as energy linearity, energy resolution, detection efficiency and temperature response can be carried out on the facilities. So far we have completed the calibration of two satellites, and there are still three satellites in progress. This work will contribute to the development of X-ray astronomy, and contribute to the development of Chinese space science.

**Keywords,** Monochromatic X-rays, Monochromator, Bragg diffraction, Energy spectra, Calibration.

*Correspondence, gsm@nim.ac.cn, wujj@nim.ac.cn.

# 1 Introduction

Beginning in the 1960s, using rockets or high-altitude balloons and other vehicles, collimating scintillation crystals or gas detectors have been used to realize space X-ray detection, and start space astronomical observation research. X-ray astronomy mainly studies high-energy radiation celestial bodies and the high-energy radiation properties of celestial bodies through the observation of X-rays. The main observation objects include black holes, neutron stars and other dense celestial bodies and interstellar high-temperature hot gases. It is a branch of astronomy that studies physical processes under extreme conditions such as extremely high density, extremely strong magnetic fields, and extremely strong gravitational fields. Take the United States as an example, during 2000-2014, a total of 84 space science satellites were launched, accounting for 27.6% of the total number of satellites launched. In 2017, the first Chinese space science satellite Hard X-ray Modulation Telescope (HXMT) was successfully launched and achieved a series of important research results [1][2]. At present, China has many space science satellites under development or about to be launched, such as The Gravitational wave high-energy Electromagnetic Counterpart All-sky Monitor (GECAM)[3], Hard X-ray Imager (HXI)[4], Space Variable Objects Monitor(SVOM)[5][6], The enhanced X-ray Timing and Polarimetry mission (eXTP)[7], etc. The detectors carried by many space science satellites are the key to space observation. The key parameters such as energy resolution, detection efficiency and uniformity of these detectors must be given through ground calibration. Otherwise, the inversion from observation data to real astronomical objects cannot be realized. Therefore, it is necessary to carry out research on monochromatic X-rays calibration facilities and calibration methods of detectors.

There are four ways to generate monochromatic X-rays, radionuclides, synchrotron radiation, K fluorescence and Bragg diffraction of X-rays tube. Radioactive sources are widely used in detector calibration, but their energy values are constant, rare and uncontrollable. The synchrotron X-ray source is widely used in many fields, but it has the problems of high cost, inconvenient application and unstable. K fluorescent X-rays are generated by the primary X-rays to excite the secondary target. Due to material limitations, they are also limited to a specific energy. Based on the continuous spectrum generated by X-ray machine, monochromatic X-rays can be obtained by Bragg diffraction, and their energies are continuous and adjustable by using different crystals and change the Bragg angle. This method is convenient, effective and low-cost in the application of detector performance calibration.

Internationally, monochromatic X-ray radiation facilities based on X-ray machines include the SOLEX device from the Becquerel laboratory in France[8][11], the XACT in Italian[12], the XRCF in America[13] and the PANTER in Germany[14], etc., and the main performance comparison is shown in Tab. 1. Satellites calibrated by these facilities include BeppoSAX[15]，INTEGRAL/JEM-X[16]，AXAF[17]，Swift[18]，GBM[19][20], etc. In China, there is the HXCF[21] jointly built by the National Institute of Metrology of P.R. China and the Institute of High Energy Physics, Chinese Academy of Sciences. Its energy range is (15-168) keV. It provides on-ground calibrations on energy linearity, energy resolution, detection efficiency, energy response matrix and effective area of Chinese first astronomical satellite, HXMT [22], which laid the foundation for its scientific achievements[1][23].

Tab. 1. The emblematic monochromatic X-rays facilities over the world.

| No. | 1 | 2 | 3 | 4 | 5 | 6 |
|---|---|---|---|---|---|---|
| Facility | XACT | XRCF | SOLEX | PANTER | Ferrara | HXCF |
| Country | Italy | USA | France | Germany | Italy | China |
| Energy(keV) | 0.1-30 | 0.09-10 | 1-20 | 0.25-50 | 15-140 | 20-161 |
| Flux(cm-2s-1) | $10^5$ | - | $10^3$ | $10^4$ | - | $10^3$ |
| Spot size(φ cm) | 0.1 | 400 | 0.035 | 100 | 0.5 | 0.5 |
| Beam length(m) | 35 | 518 | 0.5 | 120 | 100 | 6 |
| Clean room | Y | Y | Y | Y | Y | N |
| Vacuum | Y | Y | Y | Y | Y | N |
| Monochromator | Optical grating | Optical grating&DCM | SCM | Optical grating&DCM | DCM | DCM |
| Monochromaticity | - | ＜1% | - | 4%@10keV | 1.7%@17keV | <1%@60keV |

# 2 The monochromatic X-ray facilities

In order to build monochromatic X-ray facilities to provide calibration and research conditions for space science satellite-borne detectors and other detectors, lots of research work had been carried out at NIM, and several facilities had been built. So far, the monochromatic X-ray facilities covering energy range of (0.218~301) keV has been completed. The energy range of (0.218~1.6) keV is generated by X-ray machines and diffracted by gratings. The (5~301) keV energy range is produced by Bragg diffraction of crystals.

## 2.1 The (0.218~1.6) keV X-ray facility

The (0.218~1.6) keV monochromatic X-rays beam facility is mainly composed of X-ray machine, non-harmonic grating monochromator, detector, etc. The bremsstrahlung radiation spectrum generated by the X-ray machine enters the non-harmonic monochromator to achieve dispersion focusing. And the adjustable monochromatic X-rays beam is obtained at the exit. The source is a customized multi-target windowless X-ray machine, from WORX, model XWT-065-SE, with five target materials. When the target materials are changed, the characteristic X-rays corresponding to different target materials can be generated. The main parameters of X-ray machine and monochromator are as shown in Tab. 2.

Tab. 2. The First Instrument Characteristics.

| X-ray Tube | | Monochromator | |
|---|---|---|---|
| Anode Voltage | 5 kV-30 kV | Energy Range | 218 eV-1600 eV; |
| Max Anode Current | 5 mA@10 kV | Incident Angle Range | 85.6º~88.6º |
| Focal Spot Size(Nominal) | 20 μm-200 μm | Monochromaticity | ΔE/E<2.5% |
| Anode Material | Si、Cu、Ti、Ag、Cr | Higher harmonic | <0.3% |
| Cooling Method | Water | Adjustable step | <10 eV |
| Flux Stability | better than ±0.5% | Vacuum degree | ≤5E-4 Pa |
| Flux linear | better than ±0.5% | Divergence angle | ≥5 mrad×10 mrad |
| Continuous working period | ≥10 h | Grating line width | 1500lp/mm |

Taking into account the convenience and the stability of the focused light source, the monochromator is directly connected to the source without a front focusing system. The non-harmonic grating monochromator adopts the CT optical structure, and uses two toroidal mirrors to realize the collimation and focusing functions. The focus positions are respectively at the entrance and exit of the slit. The incident and diffracted beams of the grating are parallel in the spectral resolution direction. When wavelength scanning, only one element of the grating needs to be rotated, and the focusing conditions can be met without moving the exit slit. The structure is shown in Fig. 1.

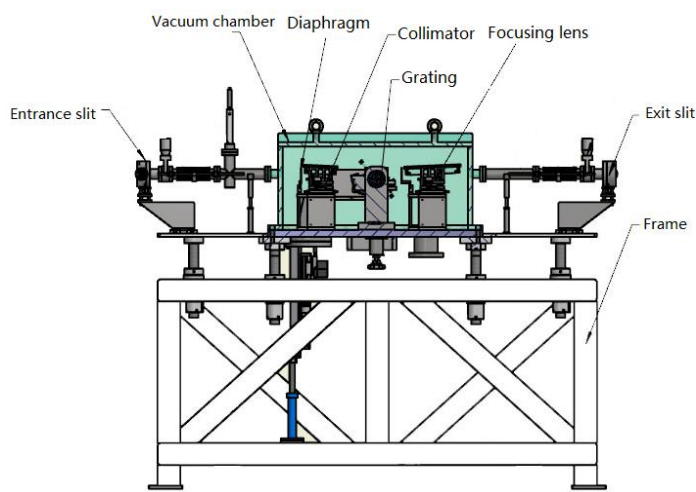

Fig. 1 Schematic configuration of Soft X-ray monochromator.

## 2.2 The (5~40) keV X-ray facility

The (5~40) keV monochromatic X-rays beam facility is mainly composed of X-ray machine, diffraction crystal and synchronous rotating device. The uncertainty of the photon flux introduced by the position uncertainty is very large. A position deviation of 1 mm may cause a 20% flux error. The position accuracy is significant for the detection efficiency calibration of the detector. The advantage of the double crystal monochromator is that there is only a very small translation in the X-ray emission direction when adjusting different energies, and no angular deflection occurs. This is beneficial to the detection efficiency calibration, and the single crystal monochromator has each energy corresponding to an angle. Every time the energy is changed, the angle needs to be adjusted, and every angle change needs to change the position of the detector, which introduces a considerable uncertainty. The best method is to make the direction of emission X-ray fixed, and there is no need to change the position of the detector when adjusting different energies. In order to ensure that the direction of emission X-ray remains unchanged, the design scheme adopted is as shown in Fig. 2. The single crystal monochromator is separately placed on a turntable. At the same time, the crystal turntable and the X-ray machine are on the same rotatable platform, and the upper and lower rotations are controlled by two motors. The upper motor controls the rotation of the crystal, and the lower motor controls the X-ray machine. When the crystal rotates at the angle $\theta$, the X-ray machine should rotate at an angle of $2\theta$ to ensure that the X-ray emission direction is consistent with that before the rotation.

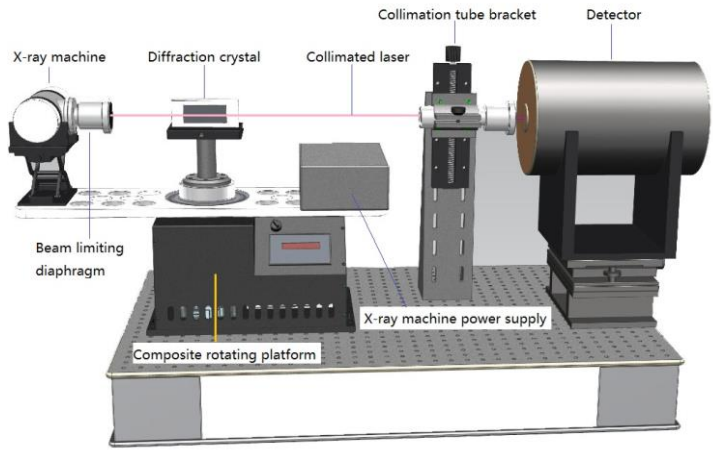

Fig. 2 Schematic configuration of the (5~40) keV monochromatic X-rays beam facility

The monochromator consists of a high-precision rotator and a Bragg diffraction crystal, the monochromatic X-rays emerging from the collimator tube pass then through the single crystal monochromator. The Bragg diffraction of X-rays can produce monochromatic X-rays, as stated by the Bragg Law [24],

$$2d \sin \theta = n\lambda \tag{1}$$

Where $d$ is lattice spacing, $n$ represents the diffraction series, $\theta$ is the angle of incidence and $\lambda$ is the wavelength of X-ray photons. From the photon energy formula, the energy of monochromatic X-ray photons can be deduced,

$$E = hv, c = \lambda v \Rightarrow E = \frac{nhc}{2d \sin \theta} \tag{2}$$

The detailed parameters are shown in tab. 3.

Tab. 3. The Second Instrument Characteristics.

| X-ray Tube | | Monochromator | |
|---|---|---|---|
| Max Anode Voltage | 15 kV-50 kV | Energy Range | 5 keV-40 keV |
| Max Anode Current | 1 mA@50 kV | Bragg Angle | 4°~25° |
| Max Filament Current | 1.7 A | Monochromaticity | ΔE/E<3%@10 keV |
| Focal Spot Size(Nominal) | 110 μm | Monochromatic light | >90% |
| Anode Material | Cu | Flux | >5000 cm$^{-1}$s$^{-1}$ |
| Be Window Thickness | 125 μm | Adjustable step | <0.2 keV |
| Stability | 0.2% over 4 hours | Spot Size | φ1~φ10 mm |
| Cooling Method | Water | Single Crystal | LiF (220), LiF(420), Ge(111) |

## 2.3 The (20~161) keV X-ray facility

The main structure of the (20~161) keV monochromatic X-rays beam facility is shown in, which is composed of an X-ray machine, a double crystal Monochromator, a standard detector and a control system. After the X-ray machine is turned on outside the shielded room, the X-ray continuous spectrum emitted by the X-ray tube passes through the front collimator and the aperture and then hits on the double crystal Monochromator. The crystal changes the angle under the control of high precision turntable and T structure, and the single energy X-rays with different energy are obtained. After passing through the collimator and the beam-limiting aperture, the monochromatic X-ray was finally detected by the detector.

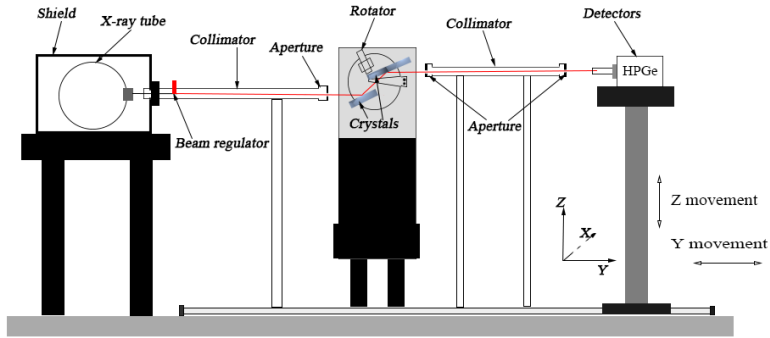

Fig. 3 Schematic configuration of the (20~161) keV monochromatic X-rays beam facility

The double crystal Monochromator device is an X-ray diffraction device with high precision and high resolution, which makes use of the Bragg diffraction law of the crystal. It includes the double crystal, the High-precision rotator, the T structure, the Beam regulator and the aperture, then the structure principle is shown in fig. 3. The monochromator uses two parallel crystals as the dispersion original, and the first crystal ("crystal I") realizes monochromaticity. Under the action of the fixed height difference structure, the second crystal ("crystal II") keeps the exit direction and height of the output monochromatic light relative to the incident light unchanged, so as to obtain a light spot with a fixed position. The detailed parameters are shown in tab. 4.

Tab. 4. The Third Instrument Characteristics.

| X-ray Tube | | Monochromator | |
|---|---|---|---|
| Anode Voltage | 10 kV-225 kV | Energy Range | 20 keV-161 keV |
| Max Anode Current | 60 mA@40 kV | Bragg Angle | 2.5º~7.5º |
| Max Filament Current | 4.2 A | Monochromaticity | $\Delta E/E<2\%$@60 keV |
| Focal Spot Size(Nominal) | 0.4 mm or 3 mm | Monochromatic light | >90% |
| Anode Material | W | Flux | >2000 $cm^{-1}s^{-1}$ |
| Be Window Thickness | 800 μm | Adjustable step | <0.2 keV |
| Stability | 0.1% over 4 hours | Spot Size | φ1~φ10 mm |
| Cooling Method | Water | Double Crystal | Si(220), Si(551) |

## 2.4 The (21~301) keV X-ray facility

The configuration of the facility is shown in Fig. 4. The facility consists of four parts, including the X-ray tube, the collimating structure, the monochromator, and the detectors. The emission X-rays from the X-ray tube goes through the collimator and aperture, then hits the surface of the crystal. The monochromatic photons are then achieved after Bragg diffraction. By means of controlling the rotator, the Bragg angle is adjustable. Different Bragg angles produce different energy values. In the end, the monochromatic light is observed by X-ray detectors. In order to accurately detect the X-ray photons, the position of the detector can be adjusted along with the change of the Bragg angle under the help of the control system.

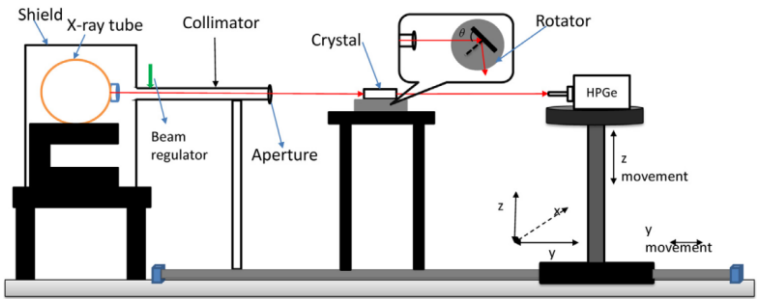

Fig. 4 Schematic configuration of the (21~301) keV monochromatic X-rays beam facility

The X-ray machine uses the Y.MG605 type X-ray machine of YXLON Company, the rated voltage range is (20-600kV) kV. There are three different silicon crystals, namely Si111, Si220, and Si551, for the Bragg diffraction. The detailed parameters are shown in Tab. 5.

Tab. 5. The Forth Instrument Characteristics.

| X-ray Tube | | Monochromator | |
|---|---|---|---|
| Max Anode Voltage | 20 kV-600 kV | Energy Range | 21 keV-301 keV |
| Max Anode Current | 7.5 mA@200kV | Bragg Angle | 1.5º~25º |
| Max Filament Current | 3.7 A | Monochromaticity | $\Delta E/E<3\%$ |
| Focal Spot Size(Nominal) | 0.5 mm or 1.5 mm | Monochromatic light | >90% |
| Anode Material | W | Flux | >2000 $cm^{-1}s^{-1}$ |
| Window Thickness | 2 mm Be and 3 mm Al | Adjustable step | <1 keV |
| Stability | 0.3% over 4 hours | Spot Size | φ1~φ10 mm |
| Cooling Method | Oil | Single Crystal | Si(220), Si(551) |

# 3 Measurements

## 3.1 Detectors

### 3.1.1 HPGe

The standard detector of is a high-purity Germanium (HPGe) Detector (Canberra GL0110). The active area is 100 $mm^2$, the crystal length is 10 mm, the Beryllium window thickness is 0.08 mm, and the resolution full width at half maximum (FWHM) at optimum settings is 160 eV (@5.9 keV) and 500 eV (@122 keV). The HPGe detector is used to record the X-ray spectrum and the photon counts. Combined with the parameters provided by the manufacturer, the internal structure of the detector is obtained by industrial CT scanning, the geometric model is established to simulate the detection efficiency by Monte Carlo codes, and the point source extrapolation experiment is verified, finally the detection efficiency of HPGe is obtained. The perspective image of HPGe is shown in Fig. 5.

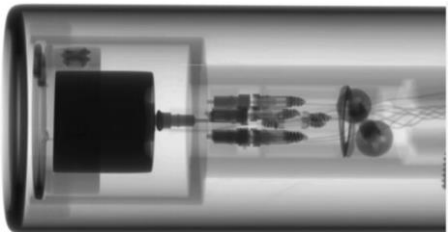

Fig. 5. X-ray radiography image of the HPGe detector.

An experimental calibration was carried out like Fig. 6. Three radioactive sources were selected in the experiment, which are $^{241}$Am, $^{57}$Co and $^{109}$Cd, after calculation of the experimental data we can get the experimental efficiencies at different energy. The specific experimental methods are given in reference [25].

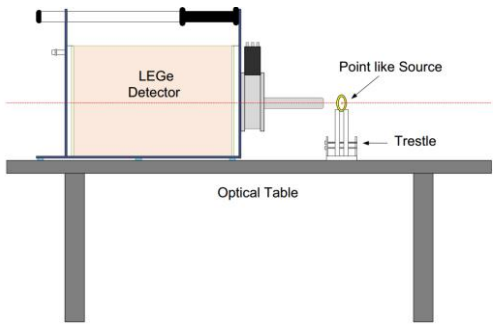

Fig. 6. Calibration experiment.

We compared the simulation and experimental efficiencies [26], after necessary corrections these results fit well. The results are listed in Tab.6.

Tab. 6. Intrinsic full-energy peak efficiency and relative deviation RD(%) between experimental and simulation methods.

| Energy(keV) | Experimental intrinsic full-energy peak efficiency (%) | Simulation Intrinsic full-energy peak efficiency (%) | RD (%) |
| --- | --- | --- | --- |
| 14.41 | 87.6(15) | 87.8(9) | -0.28 |
| 22.08 | 92.9(10) | 93.29(10) | -0.47 |
| 59.54 | 97.8(9) | 98.38(10) | -0.52 |
| 88.03 | 94.6(11) | 94.81(10) | 0.04 |
| 122.06 | 72.6(9) | 73.88(9) | -1.8 |
| 136.47 | 63.4(10) | 63.35(8) | 0.07 |

The HPGe detection efficiency curve is shown in the figure below. The test data points in the figure are measured by the experiments according to the point source extrapolation method. The curve is calculated by establishing a Monte Carlo model based on the CT scan image of the HPGe detector. This curve is used for beam flux calculation.

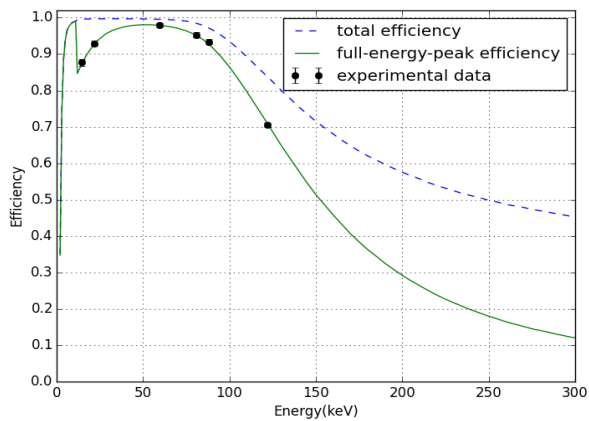

Fig. 7 The detection efficiency obtained from MC for parallel incident photons and experiment

### 3.1.2 SDD

Silicon drift detector (SDD) is another kind of semiconductor detector (AXAS-D from KETEK) commonly used in X-ray detection. Because of its counting rate, high energy resolution and the ability, it is widely used in scientific research. The description of the detector are as follows, the active area is 20 mm$^2$, the crystal length is 450 μm, the Beryllium window thickness is 8 μm, and the resolution full width at half maximum (FWHM) at optimum settings is better than 133 eV @5.9 keV.

Efficiency calibration is the experimental work to determine the efficiency of a nuclear radiation detector in recording incident particles. A common method is direct calibration using a series of standard radioactive source with known rates of decay and branching. That can also be obtained by Monte Carlo simulation and then calibrate a few points with standard radioactive sources. The simulations were performed using the Monte Carlo codes to calculate the SDD detector efficiency. A detector model is necessary for Monte Carlo simulations. The most accurate and effective method is to perform fluoroscopic imaging of the detector. We performed X-ray radiographs of the detector to check the nominal dimensions specified by the manufacturer and the X-ray radiography. The dimension and structure of the simulation model are shown in Fig. 8.

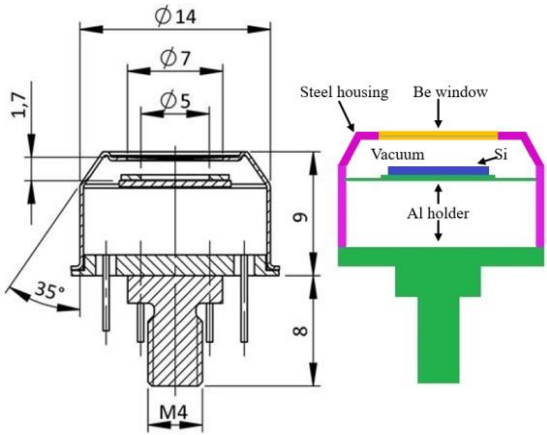

Fig. 8. Schematic representation of the dimension and configuration of the SDD.

Monte Carlo simulation results of the detector calculated by MC are shown in Fig. 5. The calibration curve contains 60 simulation results in the energy interval from 0.1keV to 40 keV. And the intrinsic efficiency of the detector varied from 2% to 99.3%. Here we only list part of the results, to be compared with later experimental ones: 99%@5.9 keV, 99%@6.5 keV, 87%@11.9 keV, 72%@14 keV, 51%@16.9 keV, 46%@17.8 keV. In addition, Fig. 2 also presents the Ge absorption edge which appears in the vicinity of 11.1 keV. It can be seen that the detection efficiency of the detector is the highest at 5-10 keV, about 99%, the photon detection efficiency drops rapidly at energies greater than 10 keV, and the detection efficiency at 40 keV is only about 5%. The reason is the crystal thickness is too small, high energy photons cannot be deposited energy within the crystal.

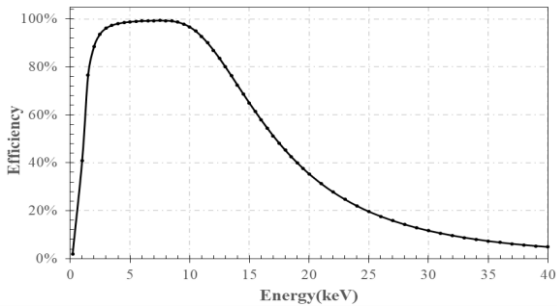

Fig. 9. Calculated intrinsic efficiecy of SDD.

### 3.1.3 CCD

We have two X-ray Charge-couple Device (CCD) cameras, one is from Princeton Instruments PIXIS-XF: 2048B. The X-ray low noise camera utilizes back illuminated CCDs, and it is designed for indirect imaging X-rays. It has high spatial resolution thanks to its 2048×2048 imaging array, and 13.5 μm×13.5 μm pixels. Dual speed operation at 100 kHz or at 2 MHz enables the camera to be used for steady state as well as for high speed application. The nonlinearity of the X-ray CCD camera is less than 1% at 100 kHz. We use it to measure the size of the X-ray light spot. The other CCD is from Andor DO934P-BN, its imaging array is 1024×1024, the pixels is 13 μm×13 μm, the maximum speed is 5 MHz. And its detection efficiency is shown in Fig.10.

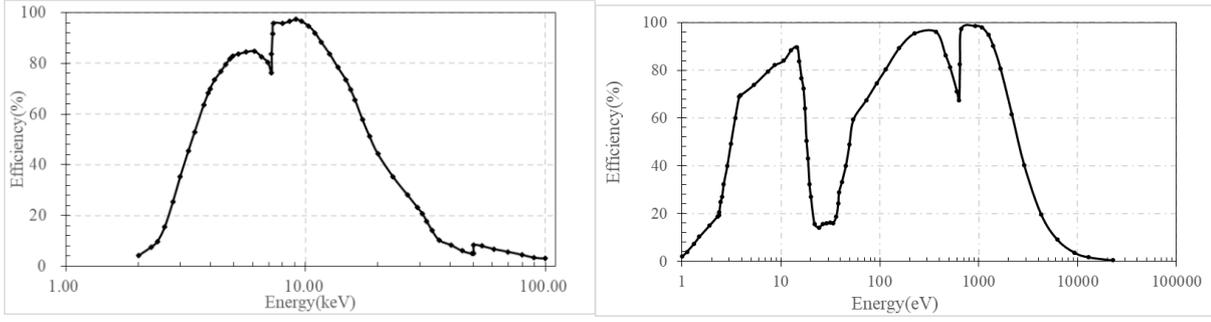

Fig. 10. Detection efficiency of PXF:2048B(left) and DO934P-BN(right).

## 3.2 Measurements and Results

### 3.2.1 Measurements of (0.218~1.6) keV photons

The zero-order spot was recorded on the CCD detector when the grating was adjusted to the horizontal position. The position coordinates of the zero-order spot's center point was (1318, 1092), the incident angle was α. The first-order spot could be recorded while the grating angle was rotated to 949.8eV (λ=1.305nm). The position of light spot center could be located in the zero-order coordinates (1318, 1092) ultimately through further rotating grating angle slightly, at this point, Δα=0.851°. According to the grating diffraction equation,

$$\sin(\alpha + \Delta\alpha) - \sin(\alpha - \Delta\alpha) = \lambda\rho \qquad (3)$$

Therefore, from Δα and λ, the actual incident angle of the zero-order α=87.0715° can be obtained. Then the first-order spot of 828eV was collected on the CCD while the grating was rotated to 0.976°, as shown in Fig. 11.

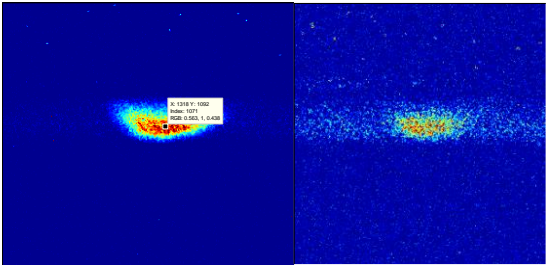

Fig. 11. Cu target X-ray source 949.8eV primary spot (left) and 828eV primary spot (right)

Replace the target with Ti, and rotate the grating to 1.762°, the first-level spot of 458.3 eV is recorded by the CCD. When the grating rotated, the limit of rotation angle is 0.5052° and 3.7°, for the 1500 lp/mm grating, the corresponding energies are 1600 eV and 218 eV respectively. Thus, the working energy range of the monochromator is 218 eV~1600 eV. On the CCD detection surface, the spot length is 103 pixels, about 1.39 mm, the spot height is 39 pixels, about 0.53 mm. The total photon counts is 1.64E6/60s, so the count rate is 2.7E4cps@949.8 eV.

At the best resolution position of the first-level spot 949.8 eV (λ=1.305nm), the grating incident angle is 87.0661°. The number of pixels for broadening of spectral lines ΔN=31(FWHM) is obtained from the measured spot. And the distance between the center of CCD array and the focusing lens is L=800 mm, hence the angle broadening corresponding to the FWHM of the spectral line is calculated as ΔN*13.5(μm)/L=5.23E-4. According to the grating equation,

$$\sin(\alpha) - \sin(\beta \pm 13.5\Delta N/(2L)) = (\lambda \mp \Delta\lambda_{\mp})/d \qquad (4)$$

We could get the energy broadening corresponding to angular broadening ΔE=20.496 eV. Furthermore,

energy resolution is obtained as E/ΔE≈46@949.8 eV (2.2%@949.8keV).

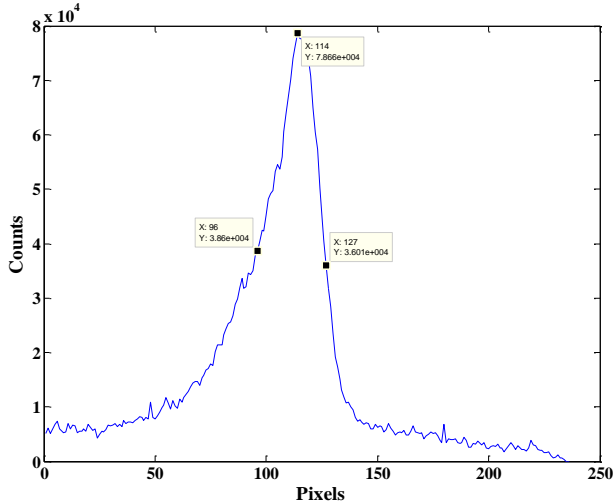

Fig. 12. FWHM of the 949.8 eV X-ray.

**3.2.2 Measurements of (5~40) keV X-rays**

The SDD detector is used to collect data every 30 minutes, and the energy stability is obtained through long-term measurement, thereby obtaining the stability of the device, it is shown in Fig. 13.

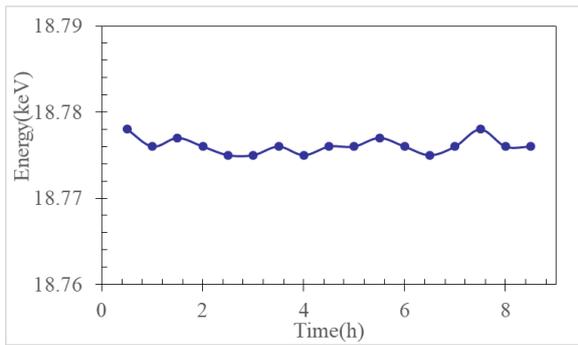

Fig. 13. The energy stability

The energy stability of the facility is 0.02%@8 h, and its flux stability is within 1.5%@8 h.

The energy range of the facility is determined experimentally, and the energy spectra are obtained by SDD and HPGe detectors. Some of the experimental results are listed in Fig. 14.

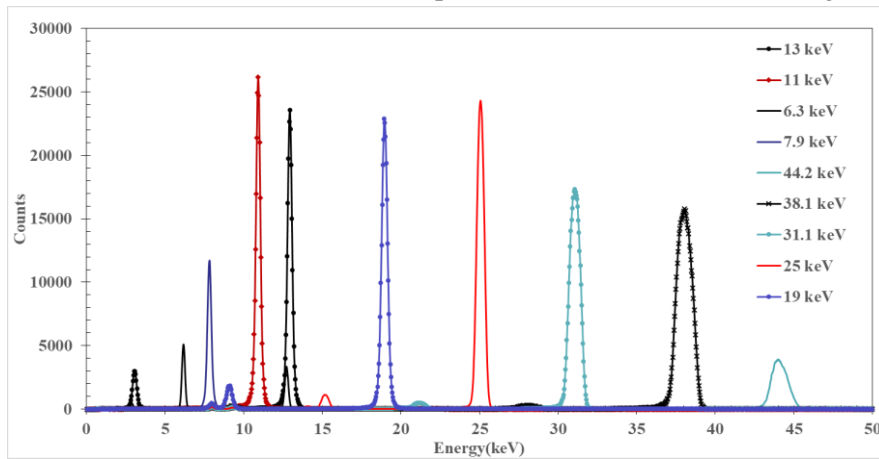

Fig. 14. The energy spectra of the monochromatic light using the LiF220 single crystal monochromator. The energy value in the title corresponds to the one recorded during the experiment.

Photons will be scattered after passing through a 3 mm beam limiting diaphragm, resulting in a spot size larger than the aperture of the diaphragm. Move the CCD detector to the center of the laser positioning. The coordinates of the edge of the spot are measured as (739,713), (1016, 713), (874,551), (874,881). The image obtained by CDD measurement is 13.5 μm per pixel, so the horizontal length of the spot is 3739.5 μm and the vertical length is 4455 μm.

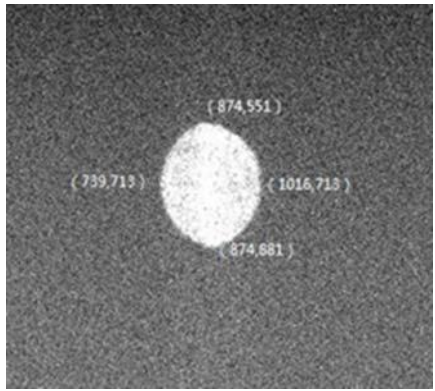

Fig. 15. Measurement spot distribution.

The HPGe detector was used to measure the performance of the device in more detail. The performance parameters such as flux and monochromaticity are listed in Tab. 7. In the current experimental mode, the minimum flux is greater than 5000 $cm^{-1}s^{-1}$. This value can be changed by adjusting the parameters of the X-ray machine. The monochromaticity is obtained by subtracting the energy resolution of the detector from the measurement result. The monochromaticity of the device in the measured energy range is below 3.24%, which is very good and can meet the calibration of almost all detectors.

Tab. 7. The measurement results from experiment.

| Energy(keV) | Count rate(cps) | Efficiency | Flux($cm^{-1}s^{-1}$) | FWHM(keV) | Monochromaticity |
|---|---|---|---|---|---|
| 6.3 | 353.29 | 0.968 | 5165.87 | 0.232 | 2.10% |
| 7.1 | 546.31 | 0.979 | 7898.49 | 0.292 | 3.11% |
| 7.9 | 956.86 | 0.981 | 13805.98 | 0.295 | 2.84% |
| 9 | 1685.76 | 0.986 | 24199.52 | 0.304 | 2.62% |
| 10 | 1905.9 | 0.983 | 27443.18 | 0.306 | 2.38% |
| 11 | 2512.4 | 0.9897 | 35931.31 | 0.327 | 2.40% |
| 12 | 2193.51 | 0.8557 | 36283.23 | 0.331 | 2.24% |
| 13 | 2558.73 | 0.8657 | 41835.49 | 0.337 | 2.12% |
| 14 | 2725.41 | 0.8752 | 44077.03 | 0.352 | 2.10% |
| 15 | 1986.58 | 0.8841 | 31804.79 | 0.365 | 2.06% |
| 16 | 1806.21 | 0.8926 | 28641.73 | 0.379 | 2.03% |
| 19 | 2964.61 | 0.9151 | 45855.02 | 0.43 | 2.01% |
| 22.1 | 2836.63 | 0.9329 | 43038.33 | 0.46 | 1.88% |
| 25.1 | 3349.95 | 0.9466 | 50091.00 | 0.554 | 2.06% |
| 28 | 3414.63 | 0.9568 | 50513.83 | 0.667 | 2.27% |
| 31.1 | 3500.82 | 0.9646 | 51370.09 | 0.83 | 2.59% |
| 34.1 | 3509.86 | 0.9705 | 51189.64 | 0.99 | 2.84% |
| 37.1 | 3073.89 | 0.9735 | 44693.07 | 1.149 | 3.05% |
| 38.1 | 4555.51 | 0.976 | 66065.54 | 1.212 | 3.14% |
| 39 | 4105.2 | 0.977 | 59474.06 | 1.266 | 3.20% |
| 40.1 | 3721.47 | 0.978 | 53859.65 | 1.315 | 3.24% |
| 42.1 | 2400 | 0.9796 | 34677.70 | 1.318 | 3.09% |
| 44.2 | 1182.65 | 0.9809 | 17065.51 | 1.221 | 2.72% |

**3.2.3 Measurements of (20~161) keV X-rays**

In this paper, the double crystal Monochromator can cover the energy range of (20-161) keV with Bragg diffraction crystal Si220 and Si551. Some spectra measured by HPGe is selected, and the double crystal Monochromator is rotated so that the maximum diffraction angle can reach the lower limit of the energy, and the minimum diffraction angle can get the maximum energy. In the experiment, due to the limitation of the structure, the rotation range of double crystal Monochromator for different crystals will be slightly different, and the energy will also be limited due to the influence of crystal length and the distance between the two crystals. In theory, various crystals have a very wide energy range, but due to the limitation of the processing structure, the theoretical range will be reduced. Each crystal has a better flux and energy resolution in the appropriate energy range. Some measured spectra are shown in Fig. 16.

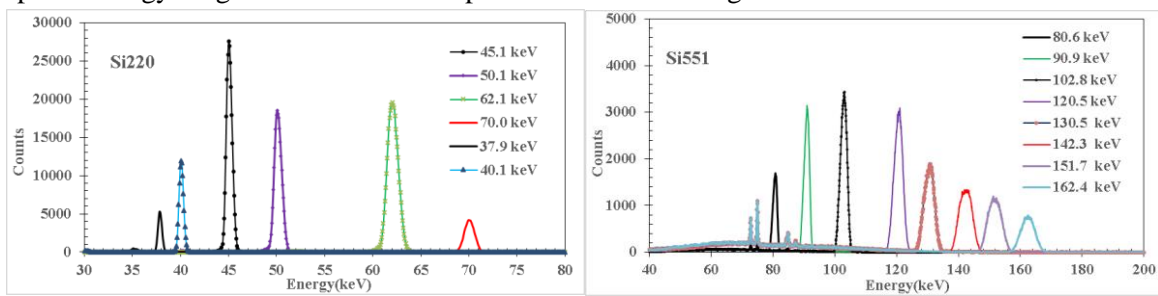

Fig. 16. The energy spectra of the monochromatic X-rays produced by the double crystal monochromator.

The flux stability is measured. After the X-ray tube is warmed up, the tube voltage and tube current are fixed, and the flux changes within 10 hours are recorded, record two sets per hour, each set for 1000s. The flux stability is better than 0.8% over 10 hours.

Under normal circumstances, a uniform spot with good monochromaticity can be obtained under 10mm. The size of the light plate mainly depends on the size of the beam limiting diaphragm. The CCD detector was used to measure the light spot under a beam limiting diaphragm with a diameter of 4 mm, as shown in Fig. 17.

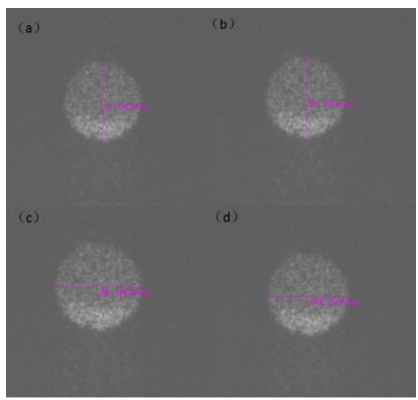

Fig. 17. Measurement of light spot.

The HPGe detector was used to measure the performance of the device in more detail. The performance parameters such as flux and monochromaticity are listed in Tab. 8. In the current experimental mode, the minimum flux is greater than 2000 $cm^{-1}s^{-1}$. The monochromaticity is obtained by subtracting the energy resolution of the detector from the measurement result. The monochromaticity of the device in the measured energy range is below 3.55%.

Tab. 8. The specification of the double crystals monochromator.

| Energy(keV) | Count rate(cps) | Efficiency | Flux($cm^{-1}s^{-1}$) | FWHM(keV) | Monochromaticity |
|---|---|---|---|---|---|
| 37.9 | 344.07 | 0.97597 | 2806.86 | 0.489 | 1.06% |
| 39.1 | 602.35 | 0.97705 | 4908.43 | 0.521 | 1.12% |
| 40.1 | 872.71 | 0.97797 | 7104.85 | 0.548 | 1.16% |
| 45.1 | 2588.1 | 0.98144 | 20995.57 | 0.708 | 1.41% |
| 50.1 | 2243.61 | 0.98324 | 18167.63 | 0.909 | 1.69% |
| 62.1 | 3439.53 | 0.98374 | 27837.43 | 1.342 | 2.08% |
| 70.1 | 691.32 | 0.98116 | 5609.83 | 1.263 | 1.71% |
| 80.6 | 344.48 | 0.96899 | 2830.45 | 1.619 | 1.94% |
| 90.9 | 808.39 | 0.93955 | 6850.33 | 2.062 | 2.21% |
| 102.8 | 1116.09 | 0.87914 | 10107.68 | 2.572 | 2.46% |
| 111.5 | 946.76 | 0.82555 | 9130.76 | 2.879 | 2.54% |
| 120.5 | 1271.27 | 0.75899 | 13335.59 | 3.271 | 2.68% |
| 130.5 | 941.63 | 0.68695 | 10913.54 | 3.982 | 3.02% |
| 142.3 | 874.2 | 0.60111 | 11578.90 | 5.081 | 3.55% |
| 151.7 | 790.42 | 0.54215 | 11607.77 | 5.381 | 3.53% |
| 162.4 | 526.63 | 0.48178 | 8702.96 | 5.162 | 3.16% |

**3.2.4 Measurements of (21~301) keV X-rays**

We measured the monochromatic X-rays by the HPGe detector, and studied the device in detail. Some typical energy spectra are shown in Fig. 18. Actually, three Si crystal were used to produce the monochromatic X-rays, and the energy range was related to different crystals. In the end, we achieved (21~301) keV monochromatic X-rays on this facility.

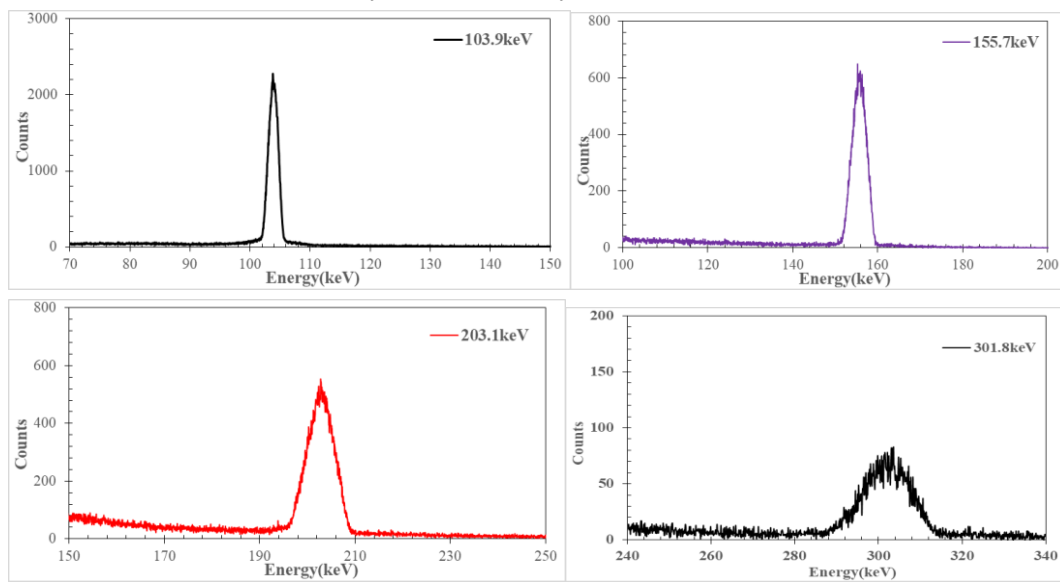

Fig. 18. The energy spectra of the monochromatic light using the Si551 single crystal monochromator.

The detailed parameters from measurement are shown in Tab. 9. Its monochromaticity is better than 5.74%, the minimum flux is greater than 2000 $cm^{-1}s^{-1}$. After detailed measurement, we found that this facility have good performance on stability and linear relationship between flux and cube current. Its energy stability is better than 0.4% within 50 $h$, the flux stability is better than 1.4% within 50 $h$. It can be used to carry out X-ray diffraction studies and detector calibration studies.

Tab. 9. The specification of the high energy monochromator.

| Energy(keV) | Count rate(cps) | Efficiency | Flux($cm^{-1}s^{-1}$) | FWHM(keV) | Monochromaticity |
|---|---|---|---|---|---|
| 51.9 | 365.5 | 0.9838 | 2957.95 | 0.565 | 0.73% |
| 93.6 | 1018.56 | 0.9292 | 8727.46 | 1.585 | 1.63% |
| 103.7 | 1079.68 | 0.8765 | 9807.39 | 1.924 | 1.80% |
| 116.8 | 1038.8 | 0.7856 | 10527.88 | 2.35 | 1.97% |
| 155.7 | 578.97 | 0.5232 | 8810.46 | 3.828 | 2.44% |
| 186.6 | 845.07 | 0.3707 | 18150.16 | 5.464 | 2.92% |
| 203.1 | 789.38 | 0.3123 | 20124.47 | 6.039 | 2.96% |
| 232.5 | 505.23 | 0.2352 | 17102.60 | 7.675 | 3.29% |
| 251.4 | 255.15 | 0.2001 | 10152.17 | 7.542 | 2.99% |
| 258.4 | 360.74 | 0.1887 | 15220.63 | 7.548 | 2.91% |
| 266.2 | 195.85 | 0.1769 | 8814.67 | 8.646 | 3.24% |
| 301.5 | 103.83 | 0.1372 | 6025.31 | 6.77 | 2.24% |
| 302.8 | 90.58 | 0.1358 | 5310.59 | 6.905 | 2.27% |
| 310.2 | 252.05 | 0.1266 | 15851.24 | 17.8 | 5.74% |

## 4. Calibration of Detectors

Recent years, several satellites dedicated to X-ray astronomy in China. The insight-HXMT (Hard X-ray Modulation Telescope) had been launched on June 15[th], 2017. The high energy X-ray telescope (HE) is one of the three instruments of HXMT payload. Its 18 main detectors and 6 backup detectors have completed the calibration of energy linearity, detection efficiency and energy resolution on our calibration device, and the energy range covers (20~150) keV [27]. The experimental energy resolution of HED Z01-25 detector is shown in Fig. 19. And the calibrated detection efficiency of all the 24 detectors on our facility are shown in Fig. 20. The experimental calibration results verify the detection efficiency of the theoretical calculation.

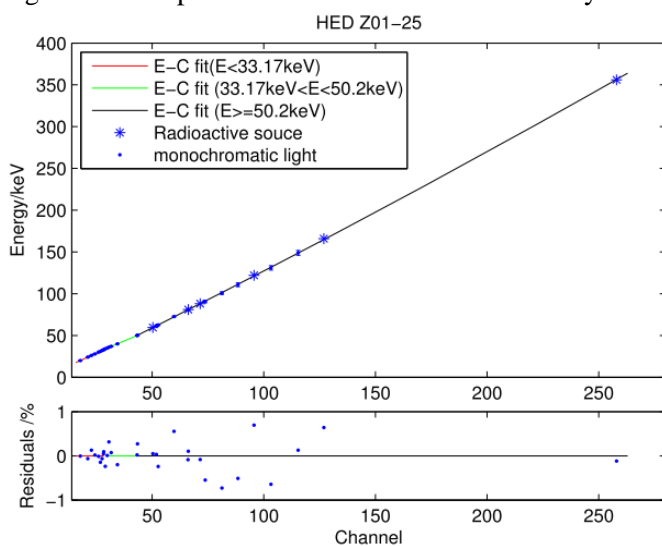

Fig. 19. Channel-energy relations for HED/NaI Z01-25.

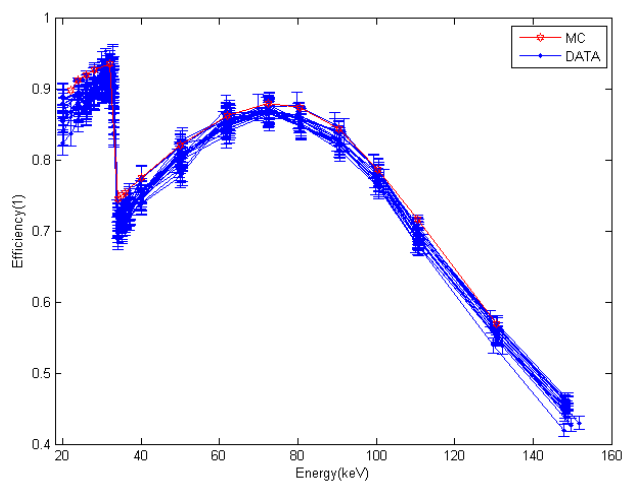

Fig. 20. Detection efficiency calibration result and MC calculation result.
The Gravitational wave high-energy Electromagnetic Counterpart All-sky Monitor (GECAM) is planned to be launched in 2020. Each GECAM satellite detects and localizes Gamma-Ray Bursts using 25 compact and novel Gamma-Ray Detectors (GRDs) in 6 keV~5 MeV. The core of the GRDs is $LaBr_3$ crystal. So far, 67 detectors (including 50 main detectors and 17 backup detectors) have been calibrated on our facilities. The energy range covers (6~160) keV. Four of the detectors have completed the fine calibration of the absorption edge (energy interval 0.1keV). The calibrating picture and some experimental results are shown in Fig. 21. and Fig. 22.

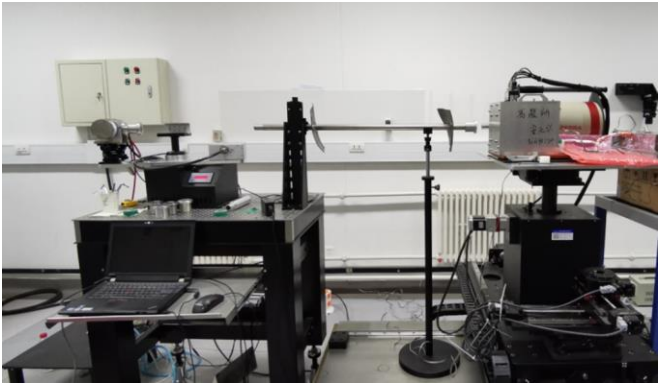

Fig. 21. Experimental picture of GRD.

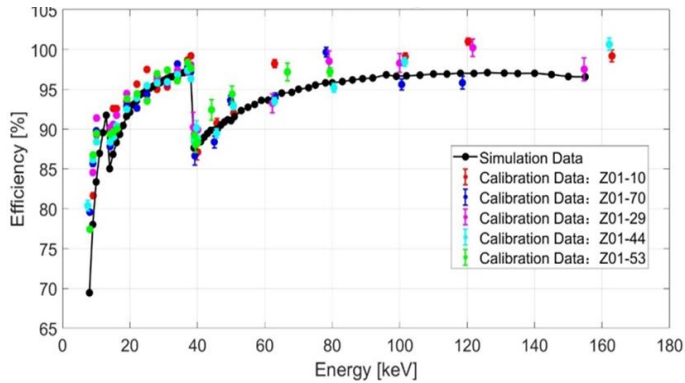

Fig. 22. Calibrated detection efficiency of some GRD detectors and MC simulation results.

In addition to the two satellite projects described above, the HXI, SVOM and Gamma Ray Integrated Detectors (GRID) will be calibrated soon. These projects have also undergone some preliminary calibration experiments.

We have also carried out some detector research work on these facilities, such as the calibration of CdTe detector X-ray CCD detector.

## 5. Conclusion

The monochromatic X-rays facilities based on X-ray tube and diffraction were introduced, and the detailed parameters were measured in the work. The standard detectors were built by Monte Carlo simulation and experimental calibration with radionuclides. The (0.218~1.6) keV monochromatic X-rays beam facility was realized in a vacuum environment. Monochromatic X-rays with good monochromaticity and stability in the measurable energy range. Limited by the structure of the device, the energy range is still relatively narrow. Improved grating support structure or use Bragg diffraction crystal to replace grating can achieve (1-10) keV monochromatic X-rays. For (5-301) keV monochromatic X-rays facilities, we have carried out a number of measurement and improvement studies to provide stable detector testing and calibration services. The measurement results show that there is a good linear relationship between flux and cube current. The experimental results also suggest that the performance of the facilities was stable.

In the future, we will broaden the energy range to higher energy and lower energy, carry out more detailed research on the monochromatic X-rays facilities below 5 keV to improve the calibration capability of this energy range. And we will try to carry out research on new standard detectors such as XTES to create conditions for providing better calibration services. We are confident to build an internationally renowned monochromatic X-ray calibration base.

## Acknowledgments

This work was supported by National Key R&D Plan of China under Grant No. 2016YFF0200802, Research Fund for the Research on Spatial Astronomical Observation X-ray Measurement Standards and Traceability Technology.